\documentclass[prl, twocolumn, superscriptaddress, showpacs,floatfix]{revtex4}  
\usepackage{hyperref}
\usepackage{amssymb}
\usepackage{amsmath}
\usepackage{float}
\usepackage{graphicx}
\usepackage{epsfig}
\usepackage{epstopdf}
\usepackage[usenames]{color}
\usepackage[normalem]{ulem}

\begin{document}

\title{Dynamical Mean-Field Theory for Quantum Chemistry} 

\author{Nan Lin}
\affiliation{Department of Physics, Columbia University, 538 West 120th Street, New York, New York 10027, USA}

\author{C. A.  Marianetti }
\affiliation{Department of Applied Physics,Columbia University, New
  York, New York 10027, USA }

\author{Andrew J. Millis}
\affiliation{Department of Physics, Columbia University, 538 West 120th Street, New York, New York 10027, USA}

\author{David R. Reichman}
\affiliation{Department of Chemistry, Columbia University, 3000 Broadway, New York, New York 10027, USA}

\date{\today}

\begin{abstract}
The dynamical mean-field concept of approximating an unsolvable many-body problem in terms of the solution of an auxiliary quantum impurity problem, introduced to study bulk materials with a continuous energy spectrum, is here extended to molecules, i.e.  finite systems with a discrete energy spectrum. The application  to   small clusters of hydrogen atoms yields ground state energies which are competitive with leading quantum chemical approaches at intermediate and large interatomic distances as well as good approximations to the excitation spectrum. 
\end{abstract}


\maketitle
Determining the ground state energy and dynamical properties of a system of $N$ interacting electrons is a fundamental unsolved problem. The properties of systems with ``strong correlations,'' those dominated by local interaction effects,  have been proven particularly difficult to treat.  Over the years many approaches have been developed \cite{Alder80,White93,Bartlett07,Hohenberg64,Kohn65,White99,Chan02,Barthel09,Verstraete08,Tsuchimochi09,Bajdich10,Hybertsen86,Holzmann09,Sharma08,Anisimov91}, but no one method has emerged as generally applicable.  In particular, for all but the smallest molecules  the treatment of correlation effects in quantum chemistry (for  example, bond breaking, or the energetics, dynamics, and magnetic properties of transition metal clusters) remains a frontier research area. 

In recent years the theoretical study of strongly correlated condensed matter systems has been revolutionized by the development of dynamical mean-field theory (DMFT), first in its single-site \cite{Georges92b,Georges96} and then in its cluster \cite{Hettler00,Kotliar01,Maier05} forms. DMFT is a Green's function based method, in contrast to many of the quantum chemical methods which are wave function based, and is often presented in terms of an impurity self-consistently coupled to a noninteracting bath of states.   To date,  the main applications \cite{Georges96,Maier05,Kotliar06} of DMFT  have been to extended (typically periodic although an interesting recent application to a nanoscopic conductor should be noted \cite{Jacobs10}) systems, characterized  by a continuous density of states. Here, we show that DMFT can be used for finite systems, for which the ``bath'' is characterized by a discrete (even small)  density of states. Application to a benchmark quantum chemical system (${\rm H}_n$, the $n$-hydrogen molecule in various configurations)  suggests the method may be useful for treating the strong correlation problems of quantum chemistry.

In the same way that  density functional theory is derived from the Hohenberg-Kohn density functional \cite{Kohn65}, DMFT may be  derived from the  Luttinger-Ward functional $\Phi_{\text{LW}}$ \cite{Luttinger60}, which is a functional of the electron Green's function $G$, the interparticle interaction ${\hat {\mathbf I}}$ and an external potential $V({\mathbf r})$. $G$ is defined in terms of the operator $\psi_\sigma({\mathbf r},t)$ which annihilates an electron at position ${\mathbf r}$ and time $t$ as $G_{\sigma\sigma{'}}({\mathbf r},{\mathbf r}{'},t-t{'})=\left<T_t\left[\psi_\sigma({\mathbf r},t),\psi_{\sigma{'}}^\dagger({\mathbf r}{'},t{'})\right]\right>$. The appropriate $G$ for a given external potential $V$ satisfies $\delta \Phi_{\text{LW}}/\delta G =0$. 

In analogy to density functional theory,  $\Phi_{\text{LW}}$ may be  written as the sum of two terms: a ``universal'' term $\Phi_{\text{LW}}^{\text{univ}}$ defined in terms of the sum of all vacuum to vacuum Feynman diagrams which has explicit dependence only on $G$ and $\hat {\mathbf{I}}$ (not on $V$) and a material-specific term which depends explicitly on $V$ and $G$ but not ${\hat {\mathbf I}}$. 
DMFT is  an exact extremization of an approximation to $\Phi_{\text{LW}}^{\text{univ}}$, in the same way that practical implementations of density functional theory are exact extremizations of approximations (such as the local density approximation) to the exact density functional.

In practice most correlated electron calculations proceed by reducing the full problem (which may be viewed as a matrix in the space spanned by the complete set of states $\phi_a({\mathbf r})$) to   a ``correlated'' subspace spanned by a set of  correlated states, $\phi^{\rm corr}_a({\mathbf r})$.  One  defines a correlated problem by retaining  only the matrix elements of ${\mathbf G}$ and $\hat{\mathbf{I}}$ within the correlated subspace and  writing a  Luttinger-Ward functional for the correlated degrees of freedom as\begin{equation}
\Phi_{\text{LW}}^{\text{corr}}=\Phi^{\text{corr}}_{\text{univ}}-\text{Tr}\ln\left[{\mathbf G}_0^{-1}{\mathbf G}_{\text{corr}}\right]+\text{Tr}\left[{\mathbf G}_0^{-1}{\mathbf G}_{\text{corr}}\right]
\label{corr}
\end{equation}
$\mathbf{G}_0\equiv \left(i\partial_t\mathbf{1}-\mathbf{H}_{\text{eff}}^{\text{corr}}\right)^{-1}$ is the noninteracting Green's function defined in the usual way in terms of the $\phi_a$ and restricted to the subspace of correlated states while $\Phi^{\text{corr}}_{\text{univ}}$ is formally defined as the sum of all vacuum to vacuum diagrams (with appropriate symmetry factors) involving ${\mathbf G}_{\text{corr}}$ and interactions $\hat{\mathbf I}^{\text{corr}}$.  

Within  the correlated subspace we define the dynamical self-energy ${\mathbf \Sigma}={\mathbf G}_0^{-1}-{\mathbf G}_{\text{corr}}^{-1}$. If the  correlated subspace contains $M$ states (summed over atoms and orbitals), ${\mathbf \Sigma}$ may be represented as  $M(M+1)/2$ functions of frequency.  The DMFT method approximates the self-energy as a sum of a much smaller number of functions.  Different versions exist \cite{Georges92b,Hettler00,Kotliar01},  corresponding to  different approximations to ${\mathbf \Sigma}$. Each  approximation to ${\mathbf \Sigma}$ implies an approximation to the Luttinger-Ward functional, which is such that the extremization may be carried out by solving a quantum impurity model with parameters fixed by a self-consistency condition. For specifics, see Ref.~\cite{Georges96,Maier05}. The results  presented here were obtained with the cellular DMFT (CDMFT) version \cite{Kotliar01}, which has a real space interpretation naturally suited to the nontranslationally invariant problem posed by a molecule. 

In CDMFT one divides the full set of states into $P$ cells, labeled by cell index $J=1,...,P$, with each cell  containing some small number $N$ of orbitals, labeled by an orbital index $a=1,...,N$. (Note that each cell may contain more than one atom.) The total number of correlated states is $M=PN$ . The approximation is to  retain only those components $\Sigma^{(J_1a_1)(J_2a_2)}$ of ${\mathbf \Sigma}$ for which $J_1=J_2$ (i.e., the two orbitals are in the same cell) and correspondingly only the intracell terms in $\hat{\mathbf I}^{\text{corr}}$.  Defining the resulting self-energy as ${\mathbf \Sigma}_{\text{CDMFT}}$ we obtain an approximation to the Green's function:
\begin{equation}
{\mathbf G}_{\text{approx}}=\left({\mathbf G}_0^{-1}-{\mathbf \Sigma}_{\text{CDMFT}}\right)^{-1}.
\label{Gloc}
\end{equation}
An approximate Luttinger-Ward functional is constructed by using ${\mathbf G}_{\text{approx}}$ in the ${\rm Tr}$ and ${\rm Tr ln}$ terms in Eq.~(\ref{corr}) and approximating the universal part as the sum of $P$ $N$-orbital quantum impurity models (different in each cell in nonperiodic cases) defined in terms of a local Green's function and the intracell portions of the interaction $\hat{\mathbf I}^{\rm corr}$. The  ${\mathbf \Sigma}_{\text{CDMFT}}^J$ are found from the solution of the  quantum impurity model for cell $J$. The stationarity condition on the approximate Luttinger-Ward functional is that the quantum impurity model Green's function for cell $J$ equals the $J-J$ component of ${\mathbf G}_{\text{approx}}$ [Eq.~(\ref{Gloc})]; enforcing this condition fixes the parameters of the quantum impurity model \cite{Georges96}.   The procedure becomes exact as cluster size $N$ approaches system size  $M$ and provides a reasonable approximation for the small values of $N$ which are computationally accessible \cite{Koch08}. Nothing in this derivation requires that ${\mathbf G}_0$ arise from a system with $P=\infty$ and a continuous density of states.

We apply the CDMFT approximation to a standard quantum chemical reference system, the ${\rm H}_n$ molecule consisting of $n$ hydrogen atoms which was recently studied by Tsuchimochi and Scuseria \cite{Tsuchimochi09} using a constrained-pairing mean-field theory (CPMFT). We  present here results for ${\rm H}_n$ chains and rings with $n=6$ and $n=50$, as well as the   ${\rm H}_4$ tetrahedron. By chain we mean a linear arrangement of atoms with open boundary conditions and interatomic spacing $R$, and by ring we mean a circular arrangement with chord distance $R$ between nearest neighbor atoms. We follow Ref. \cite{Tsuchimochi09} and define the correlated subspace as the set of hydrogen $1s$ orbitals (with both spin directions) centered on each of the hydrogen atom positions ${\vec R}_a$:  $\phi_{a,\sigma}=\psi_{1s}({\vec R}-{\vec R}_a)$. The interaction ${\hat{\mathbf I}}^{\text{corr}}$  is obtained from the appropriately antisymmetrized Coulomb integrals, which we compute using the standard minimal STO-6G basis \cite{Hehre69}.   The long range of the interaction means that all sites are coupled. In practice one can treat accurately only a subset of the interactions; the others must be treated  more approximately. We present energies in standard atomic units (a.u.); in the STO-6G basis the energy of the isolated H atom is -0.471 a.u. We use unrestricted Hartree-Fock (UHF); other choices (e.g., density functional theory) are also possible \cite{Kotliar06}.

\begin{figure}[t]
\begin{center}
\includegraphics[width=0.8\columnwidth,angle=0]{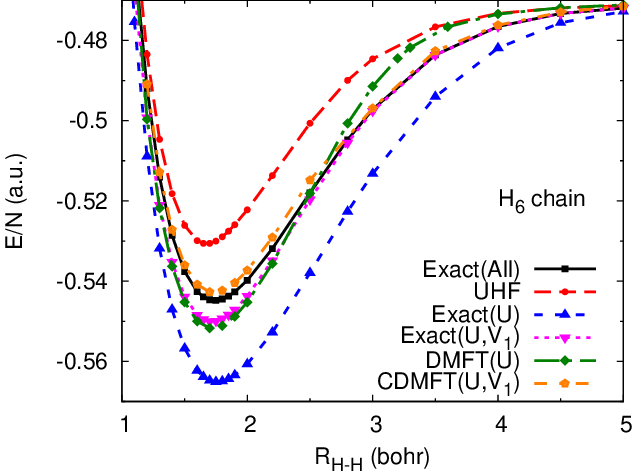}
\includegraphics[width=0.8\columnwidth,angle=0]{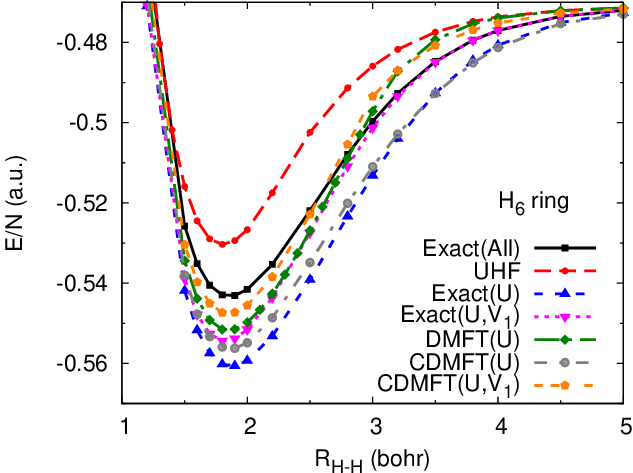}
\caption{Energy of ${\rm H}_6$ chain (upper panel) and ring (lower panel) as a function of interatomic spacing $R$ calculated with different methods as indicated.}\label{H6energies} 
\end{center}
\end{figure}

To determine which interactions must be retained we observe that for  $n=6$ the problem posed by the correlated ($1s$) subspace with the complete interaction can be  diagonalized exactly. The upper panel of Fig.~\ref{H6energies} compares the resulting energies to those obtained by treating some parts of the interaction via UHF and the other parts exactly. We focus on the  on-site term, conventionally denoted $U={\hat{\mathbf I}}^{\text{corr}}_{a,a,a,a}$ and the first neighbor terms ${\hat{\mathbf I}}^{\text{corr}}_{a,a,a,b},{\hat{\mathbf I}}^{\text{corr}}_{a,a,b,b},{\hat{\mathbf I}}^{\text{corr}}_{a,b,a,b}$ with $b$ a nearest neighbor of site $a$, which we lump together into a term $V_1$.  We define two models, ``Exact($U)$,'' where we treat the $U$ term exactly and  all other interactions  by UHF, and ``Exact($U,V_1)$'' where we treat the $U$ and all $V_1$ terms exactly and all other interactions by UHF.  
From Fig. ~\ref{H6energies} we see that  the  Exact($U,V_1$) approximation is  much closer to the exact energy than the Exact($U$) approximation. We conclude that it is important to use a method which incorporates the on-site and first neighbor interactions, while the remaining terms may be treated approximately. 

Each site has two (counting spin degeneracy) orbitals, so as defined above the CDMFT method with $N=2$ means the cluster contains one atom. At this level we have  an approximation to the Exact($U$) Hamiltonian which we refer to as ``DMFT($U$).'' The CDMFT method with $N=4$ corresponds to a two-atom cluster and constitutes  an approximation to the $\text{Exact}(U,V_1)$ Hamiltonian which we refer to as  ``CDMFT($U,V_1$).'' We have also used CDMFT as   an approximation to the Exact($U$) Hamiltonian.   We refer to this approximation as ``CDMFT($U$).''

\begin{figure}[b]
\begin{center}
\includegraphics[width=0.8\columnwidth,angle=0]{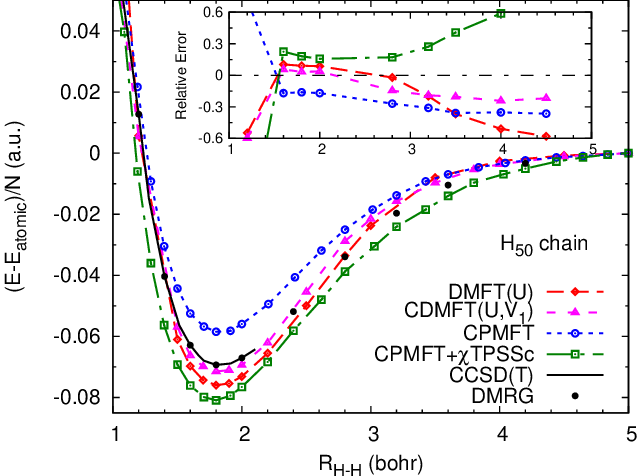}
\caption{Main panel: Energy of $\text{H}_{50}$ chain as a function of interatomic separation $R$  calculated by the   methods of this Letter and compared to digitization of data presented in Ref.~\cite{Tsuchimochi09} for coupled cluster methods (CCSD), DMRG (Ref.~\cite{Hachmann06}) and other wave function based methods (CPMFT). Inset: Relative error of CPMFT and DMFT methods, defined as $(E_{\text{method}}-E_{\text{DMRG}})/(E_{\text{DMRG}}-E_{\text{isolated atom}})$. }\label{H50energies}
\end{center}
\end{figure}

We have used these methods to calculate the ground state energies of the ${\rm H}_6$ chain  and ring.  In most of our calculations we have used the ``exact diagonalization'' (ED) method \cite{Caffarel94}  to solve the quantum impurity model. Up to $10$ bath sites were used in the CDMFT($U,V_1$) calculation of the ${\rm H}_{50}$ chain;  convergence with a number of bath sites was verified. In a few  cases we also verified that a continuous-time quantum Monte Carlo (CT-QMC) method \cite{Werner06} gives identical results.  The two panels of Fig.~\ref{H6energies} show that each approximation produces a result which lies somewhat above the exact result for the Hamiltonian which it approximates, with CDMFT providing a better approximation than DMFT.  Also, although it is difficult to perceive in the figures, the DMFT($U$) equations  have a phase transition to an antiferromagnetic state at $R\sim 4$ bohr while the Exact($U,V_1$) does not have any phase transition.  Both DMFT approximations locate the minimum in the $E(R)$ curve at essentially the exact position ($\sim 1.75\ \text{bohr}$), unlike the UHF approximation ($\sim 1.7\ \text{bohr}$).  A recent Hubbard model study \cite{Koch08} found that going from single-site to two-site clusters improved the energy substantially;  larger cluster sizes converged slowly to the exact result and the differences between two-site and exact results are less than the errors involved in constructing the approximate Hamiltonians, and will not be of interest here. 

\begin{figure}[b]
\begin{center}
\includegraphics[width=0.8\columnwidth,angle=0]{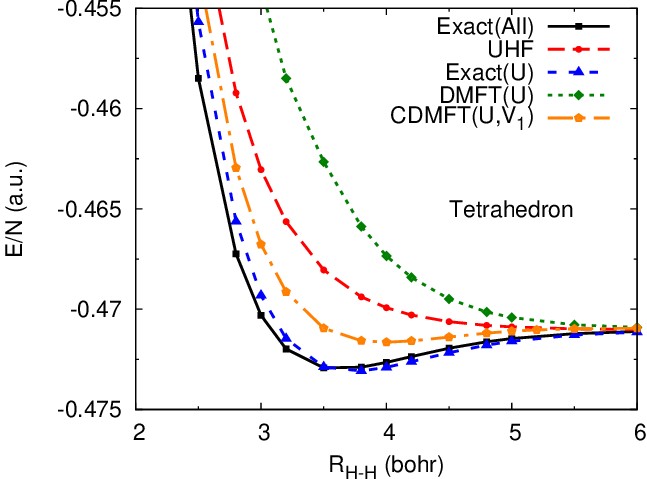}
\caption{Energy curves of tetrahedrally configured ${\rm H}_4$ calculated by different methods as indicated.}\label{tetrahedron} 
\end{center}
\end{figure}

The main panel and inset of Fig.~\ref{H50energies} compare the energies of a $50$ atom hydrogen chain obtained by our single-site and two-site cluster DMFT methods to the energies obtained from other approximate methods \cite{Tsuchimochi09} and from the  density matrix renormalization group (DMRG), which is believed to provide a numerically exact solution to the problem. The solid  line is a coupled cluster calculation also reported in Ref.~\cite{Tsuchimochi09}; this weak coupling method fails to converge as the separation increases beyond $R>2.1$ bohr.  The dotted and dashed lines labeled CPMFT are obtained from a variational wave function method \cite{Tsuchimochi09}. The inset, which shows errors relative to DMRG, demonstrates  that even the relatively primitive DMFT methods used here are generically accurate over the whole intermediate to strongly correlated regime ($R\geq 1.5$ bohr).  In the weakly correlated $(R<1.5$ bohr) regime, other methods (for example coupled cluster) are preferred, but these methods fail in the strongly correlated (here  $R\geq2.1$ bohr) regime.

Figure~\ref{tetrahedron} shows energies obtained for  the tetrahedrally coordinated ${\rm H}_4$ molecule, which is methodologically challenging  because it is only weakly bound.  Here UHF fails qualitatively, predicting that the molecule is not bound at all. DMFT($U$) is qualitatively worse, even though the intersite terms in the Hamiltonian make only a small contribution to  the ground state energy, as is seen from the close correspondence of the Exact($U$) and exact energy traces. Remarkably, the CDMFT($U,V_1$) trace [applied, of course, to the $\text{Exact}(U,V_1)$ model] produces a reasonable approximation to the energy, lying much closer to the exact curve and, in particular, predicting a minimum (although not quite at the correct location) even though the UHF curve does not. This illustrates that appropriately chosen cluster methods can capture even quite subtle behavior. 

\begin{figure}[t]
\begin{center}
\includegraphics[width=0.8\columnwidth]{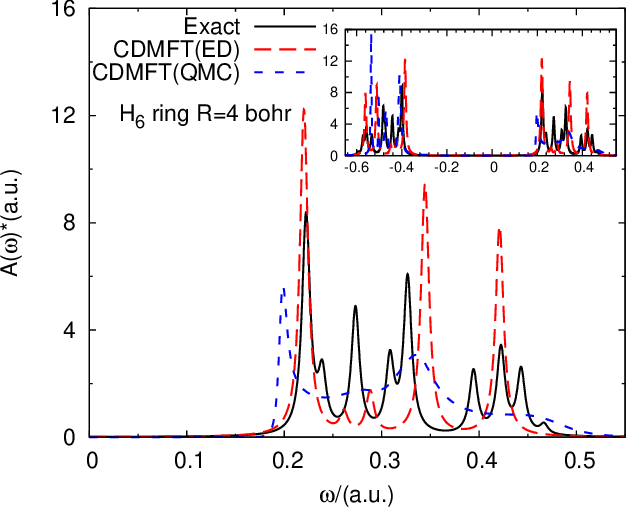}
\caption{Main panel: Electron addition spectra for ${\rm H}_6$ ring calculated from exact diagonalization of the full Hamiltonian in the correlated subspace and from the CDMFT($U,V_1$) equations using both ED and CT-QMC solvers. Inset: Full frequency range.}\label{H6spectra} 
\end{center}
\end{figure}

Figure~\ref{H6spectra} presents the electron spectral function (many-body density of states) projected onto one site of an ${\rm H}_6$ ring. The spectra are  discrete, and  the exact solution and the ED curves  have been artificially broadened. For the CT-QMC curve we have used the    maximum entropy methods of Ref.~\cite{Wang09} without any additional broadening. We see that  CDMFT($U,V_1$)  (with either solver)  provides a good approximation to the exact curve, reproducing the gap in the excitation spectrum and the basic structure of the electron addition and removal spectra.

In summary, we have shown that in the intermediate to strong correlation regime, appropriately chosen DMFT methods   give results which are superior to  other approximate methods. This, and the observation that the computational cost scales linearly with the system size (with a prefactor which depends strongly on the DMFT cluster), motivates a broader exploration of DMFT based methods in the quantum chemical context.  A more systematic comparative investigation of the merits of the different solvers available for  the quantum impurity model is needed.   The relation between cluster geometry and  molecule geometry should be more fully explored.   Density functional or GW methods may be better ways to treat the interactions not included in our DMFT calculation.  Also, alternative formulations of cluster methods better suited to finite systems may exist. 

Examination of more complicated systems, where the partitioning into correlated and ``passive'' subspaces and determination of which interactions to treat  becomes more of an  issue, is important.  More generally,  molecules provide a new context in which to examine basic theoretical issues including the ``double counting correction'' needed when combining dynamical mean-field and density functional theory, as well as the possibility \cite{Kotliar06} of using dynamical mean-field self-consistency ideas to more systematically define the correlated subspace. Research in all of these directions is in progress.

{\it Acknowledgements:} We thank Garnet Chan for helpful conversations. AJM was supported by the National Science Foundation under Grants No. DMR-075847 and No.No.  1006282 and NL  by the Nanoscale Science and Engineering Initiative of the National Science Foundation under Grant No. CHE-0641523 and by the New York State Office of Science, Technology, and Academic Research (NYSTAR). Part of this research was conducted at the Center for Nanophase Materials Sciences, which is sponsored at Oak Ridge National Laboratory by the Division of Scientific User Facilities, U.S. Department of Energy.

\bibliography{molecule_ajmjan27_cut.bbl}
\end{document}